\documentclass[12pt,twoside]{article}   
\usepackage[super,sort,comma]{natbib}

\usepackage{fancyhdr}		

\usepackage[section]{placeins}   %

\usepackage{graphicx}

\makeatletter \renewcommand\@biblabel[1]{$^{#1}$} \makeatother
\newcommand{\note}[1]{\mbox{}\\ \noindent \rule{16cm}{0.5mm} \\
{\em #1} \\ \noindent \rule{16cm}{0.5mm}
\typeout{    }
\typeout{***********note active on this page *************************}
\typeout{Note: #1  }
\typeout{****************************************end Note}
}

\usepackage[mathlines]{lineno}

\usepackage{hyperref}
\hypersetup{ colorlinks,
    citecolor=blue,
    filecolor=blue,
    linkcolor=blue,
    urlcolor=blue
}

\usepackage{xcolor}
\usepackage{amssymb}
\usepackage{lipsum}
\usepackage{placeins}
\usepackage{amsmath}
\usepackage{float}
\usepackage{hyperref}
\usepackage{algorithm}
\usepackage{algpseudocode}
\usepackage{stfloats}
\usepackage{booktabs}
\usepackage{url}

\definecolor{gray}{rgb}{0.6,0.6,0.6}
\definecolor{red}{rgb}{0.85,0,0}
\definecolor{green}{rgb}{0,0.85,0}
\definecolor{blue}{rgb}{0,0,0.85}
\definecolor{beige}{rgb}{0.92,0.87,0.78}
\usepackage[all]{hypcap}    

\begin{document}

\begin{center}
    \sf {\Large {\bfseries A Deep Learning Model for Coronary Artery Segmentation and Quantitative Stenosis Detection in Angiographic Images}} \\
    \vspace*{10mm}
    Baixiang Huang$^{1}$, Yu Luo$^{1}$, Guangyu Wei$^{2}$, Songyan He$^{1}$, Yushuang Shao$^{3}$, Xueying Zeng$^{1}$, Qing Zhang$^{4}$ \\
    $^{1}$School of Mathematical Sciences, Ocean University of China, Qingdao, 266100, China \\
    $^{2}$School of Haide, Ocean University of China, Qingdao, 266100, China \\
    $^{3}$School of Ocean and Atmosphere, Ocean University of China, Qingdao, 266100, China \\
    $^{4}$Department of Cardiology, Qilu Hospital (Qingdao), Cheeloo College of Medicine, Shandong University, Qingdao, 266035, China
\end{center}

\pagenumbering{roman}
\setcounter{page}{1}
\pagestyle{plain}

Author to whom correspondence should be addressed. Email: zxying@ouc.edu.cn (Xueying Zeng)  Email: qingzhang2019@foxmail.com (Qing Zhang)

\begin{abstract}
\noindent {\bf Background:} Coronary artery disease (CAD) is a leading cause of cardiovascular-related mortality, and accurate stenosis detection is crucial for effective clinical decision-making. Coronary angiography remains the gold standard for diagnosing CAD, but manual analysis of angiograms is prone to errors and subjectivity.\\
{\bf Purpose:} This study aims to develop a deep learning-based approach for the automatic segmentation of coronary arteries from angiographic images and the quantitative detection of stenosis, thereby improving the accuracy and efficiency of CAD diagnosis.\\
{\bf Methods:} We propose a novel deep learning-based method for the automatic segmentation of coronary arteries in angiographic images, coupled with a dynamic cohort method for stenosis detection. The segmentation model combines the MedSAM and VM-UNet architectures to achieve high-performance results. After segmentation, the vascular centerline is extracted, vessel diameter is computed, and the degree of stenosis is measured with high precision, enabling accurate identification of arterial stenosis.\\
{\bf Results:} On the mixed dataset (including the ARCADE, DCA1, and GH datasets), the model achieved an average IoU of 0.6308, with sensitivity and specificity of 0.9772 and 0.9903, respectively. On the ARCADE dataset, the average IoU was 0.6303, with sensitivity of 0.9832 and specificity of 0.9933. Additionally, the stenosis detection algorithm achieved a true positive rate (TPR) of 0.5867 and a positive predictive value (PPV) of 0.5911, demonstrating the effectiveness of our model in analyzing coronary angiography images.\\
{\bf Conclusions:} SAM-VMNet offers a promising tool for the automated segmentation and detection of coronary artery stenosis. The model's high accuracy and robustness provide significant clinical value for the early diagnosis and treatment planning of CAD. The code and examples are available at \url{https://github.com/qimingfan10/SAM-VMNet}.
\end{abstract}
\note{Baixiang Huang and Yu Luo contributed equally to this work.}

\tableofcontents

\newpage

\setlength{\baselineskip}{0.7cm}      

\pagenumbering{arabic}
\setcounter{page}{1}
\pagestyle{fancy}

\section{Introduction}
Coronary artery disease (CAD) is one of the most common cardiovascular diseases. According to the latest data from the World Health Organization (WHO), cardiovascular disease is currently the fastest-growing cause of death worldwide\cite{who2021}. This condition is characterized by the accumulation of plaque within blood vessels, a process known as atherosclerosis, which leads to vascular narrowing and hardening. This progression subsequently causes ischemic changes in tissues or organs, increasing the risk of angina, myocardial infarction, and other cardiovascular events \cite{falk2006pathogenesis}.

In clinical practice, coronary angiography is considered the "gold standard" for diagnosing CAD \cite{husmann2007coronary}. It is a non-invasive technique that examines the coronary arteries by injecting a contrast medium through the veins, allowing imaging of the heart \cite{dehkordi2019retraction}. Coronary angiography can visualize the main trunks of the left and right coronary arteries and their branches, providing information on the presence of stenotic lesions in the vessels. However, automated recognition of coronary anatomy faces several challenges: (1) overlapping structures such as catheters, the spine, and ribs; (2) low signal-to-noise ratio \cite{lin2005extraction}; (3) uneven contrast enhancement \cite{chakrabarti2014angiographic}; and (4) the subjective nature of visual interpretation in coronary angiographic images, which can lead to errors and significant variations in stenosis evaluation \cite{kim2010new}. Therefore, developing highly accurate automated methods for coronary segmentation and stenosis evaluation is of significant clinical value.

This study aims to develop an automated method for vessel contour extraction and quantitative stenosis detection. We propose a novel deep learning framework, SAM-VMNet, which combines the advantages of VM-UNet and MedSAM to achieve high-precision segmentation in coronary angiography images. The framework utilizes a transformer module to capture both global and local features and employs a selective state-space model to enhance the processing of long-sequence data. Subsequently, vessel centerlines are obtained through erosion operations, and diameter variations are measured using the maximum inscribed circle algorithm to identify and assess the severity of stenotic points. Experimental results show that SAM-VMNet outperforms VM-UNet and other advanced models in image segmentation, achieving an accuracy of 0.9772, an F1 score of 0.7736, a specificity of 0.9903, a sensitivity of 0.7409, and a mean Intersection over Union (IoU) of 0.6308. Based on clinical standards, our stenosis detection method achieved a true positive rate of 0.5867, a positive predictive value of 0.5911, an absolute root mean square error (ARMSE) of 1.7338, and a relative root mean square error (RRMSE) of 0.1673, demonstrating its potential for clinical applications.

The paper is organized as follows: In Section 2, we provide a review of related work on medical image segmentation and vascular stenosis detection methods. Section 3 describes the materials and methods used in our study, including the datasets, the proposed SAM-VMNet architecture, and the dynamic stenosis detection algorithm. In Section 4, we present the experimental results and provide a comparative analysis of our model's performance against other advanced models. Section 5 discusses the implications of our findings, clinical applications, and the limitations of our approach. Finally, in Section 6, we conclude the paper and outline potential directions for future research.

\section{Related work}
To provide context for our approach, this section reviews related work in medical image segmentation and vascular stenosis detection.  We discuss the main algorithms, highlighting both traditional methods and recent advancements in deep learning.

\subsection{Algorithms of medical image segmentation}
Traditional angiographic segmentation methods include thresholding \cite{paulinas2008algorithm}, Canny edge detection \cite{orujov2020fuzzy}, region-growing \cite{eiho2004branch}, and tracking-based approaches \cite{makowski2002two, carrillo2007recursive, manniesing2007vessel}. However, these techniques often lack robustness in challenging cases.

In recent years, deep learning methods have garnered significant attention in medical image segmentation, with Convolutional Neural Networks (CNNs) being at the forefront. The UNet architecture \cite{ronneberger2015unet} has become a staple in biomedical image segmentation, excelling in tasks such as the ISBI Neuronal Structure Segmentation Challenge. The Segment Anything Model (SAM) \cite{kirillov2023segment}, based on Vision Transformer (ViT), offers zero-shot transfer to new tasks. MedSAM \cite{ma2024segment}, trained on millions of medical images, is particularly well-suited for medical imaging applications.

The Mamba model \cite{gu2023mamba} utilizes Selective State-Space (S4) to efficiently handle long-sequence data, while VM-UNet \cite{ruan2024vm}, built on Mamba, leverages a Visual State-Space (VSS) module for contextual information, excelling in large-scale medical image segmentation. Despite these advancements, challenges persist, particularly with small or severely stenotic vessels and poor-quality images.

\begin{figure*}[t] 
\centering
\includegraphics[width=\textwidth]{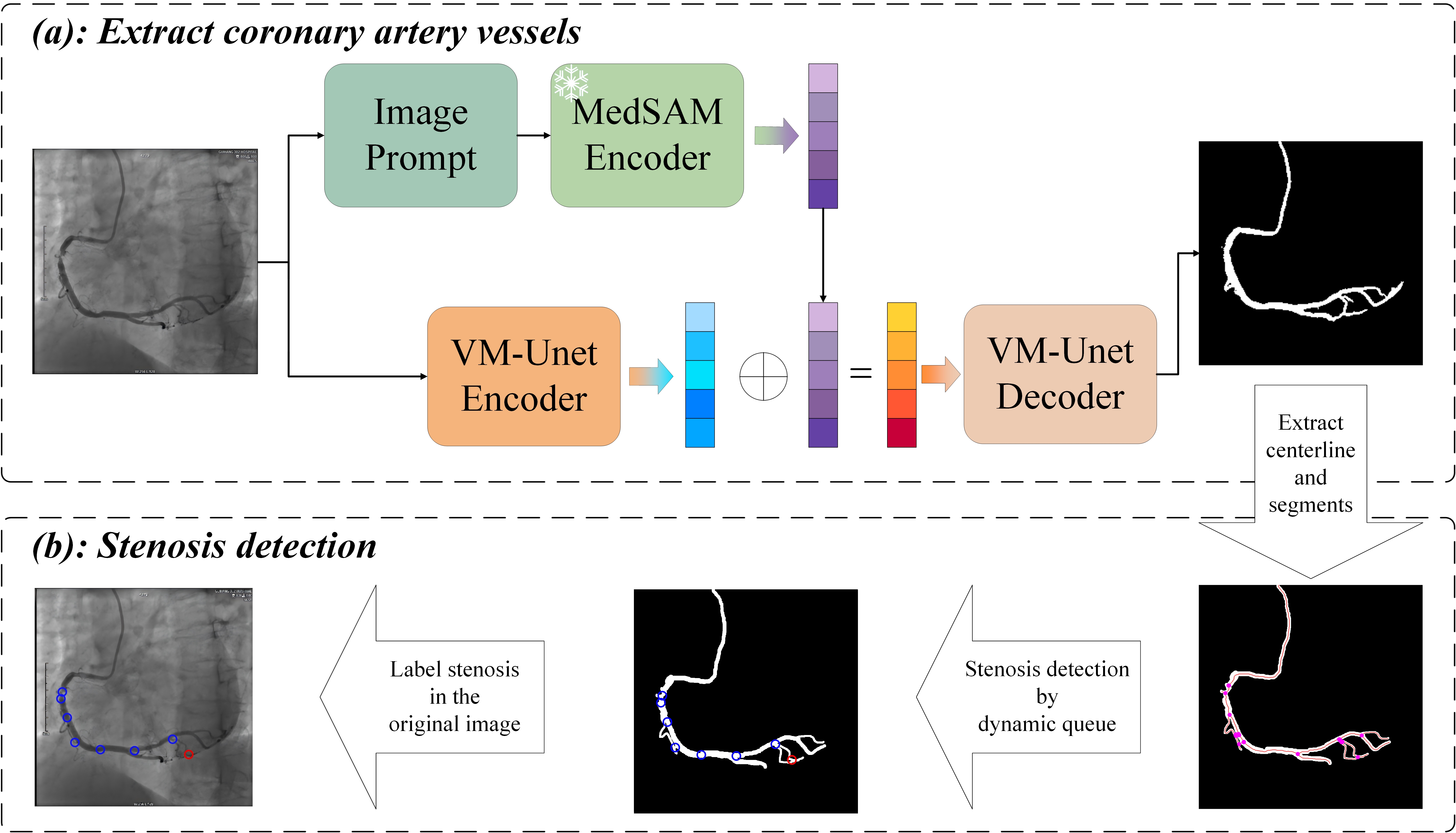}
\caption{Workflow of our method. (a) Artery segmentation using a deep learning-based approach; (b) Stenosis detection, including arterial centerline extraction, arterial diameter calculation, and stenosis degree determination.}
\label{fig1}
\end{figure*}

Recent deep learning methods for coronary angiography images have greatly improved artery segmentation. Park et al. \cite{park2023selective} proposed ensemble methods that enhance segmentation accuracy and minimize errors. Deng et al. \cite{deng2025multi} introduced MDA-Net, which uses multi-scale dual attention to capture both spatial and channel-wise information, outperforming traditional models. Zhu et al. \cite{zhu2021coronary} employed PSPNet, achieving superior accuracy compared to U-Net. Xu and Wu \cite{xu2024g2vit} developed G2ViT, which integrates CNNs, graph neural networks, and Vision Transformers to improve vessel segmentation. Lastly, Deng et al. \cite{deng2024dfa} proposed DFA-Net, which incorporates contrast-enhanced images to boost segmentation performance, particularly for small vessels.

\subsection{Algorithms for vascular stenosis detection}

The detection of vascular stenosis typically consists of three main steps: centerline extraction, diameter measurement, and stenosis degree calculation. For centerline extraction, commonly used algorithms include morphology-based methods that use erosion operations to refine blood vessels \cite{khdhair2013blood,guo1989parallel, lam1992thinning}. Another approach is based on ridge tracking, where ridges, representing grayscale extrema along the vessel direction in coronary artery images, are tracked to obtain the vessel's centerline \cite{sheng2019extraction, xiao2013automatic}. Additionally, minimum-path-based algorithms define the path with the lowest cost as the vessel centerline, with the cost function calculated using various operators \cite{lidayova2016fast}. Dijkstra's algorithm is often used for discrete $L_1$ path optimization to extract the vessel skeleton \cite{lesage2009review}.

For estimating the diameter of the vessel, the edge intersection method is a common approach \cite{zhao2010method}. This method estimates the vessel diameter by calculating the distance between intersection points on perpendicular lines to the vessel centerline. Another method, the maximum inscribed circle approach  \cite{sui2019novel}, calculates the diameter of the largest circle that can fit within the vessel at each centerline point. Additionally, the MOM method \cite{wan2018automated}, based on the topological structure of vessel images, employs a registration approach suitable for vessel tracking and measurement.

To evaluate the degree of stenosis, common calculation methods include determining the stenosis rate for each segment of the vessel by calculating the ratio of the minimum diameter to the maximum diameter within that segment \cite{liu2022two}. For each point, the stenosis rate is also calculated as the ratio of the diameter at that point on the centerline to the average diameter of the vessel segment \cite{huang2024development}. However, these methods do not account for local variations in vessel diameter, as the diameter typically narrows progressively toward the distal end. There are also object detection-based methods for detecting stenosis \cite{ovalle2022improving}, which can effectively identify stenosis but do not provide a quantitative evaluation of stenosis.

\section{Materials and methods}

\begin{figure*}[t]
    \centering
    \includegraphics[width=\textwidth]{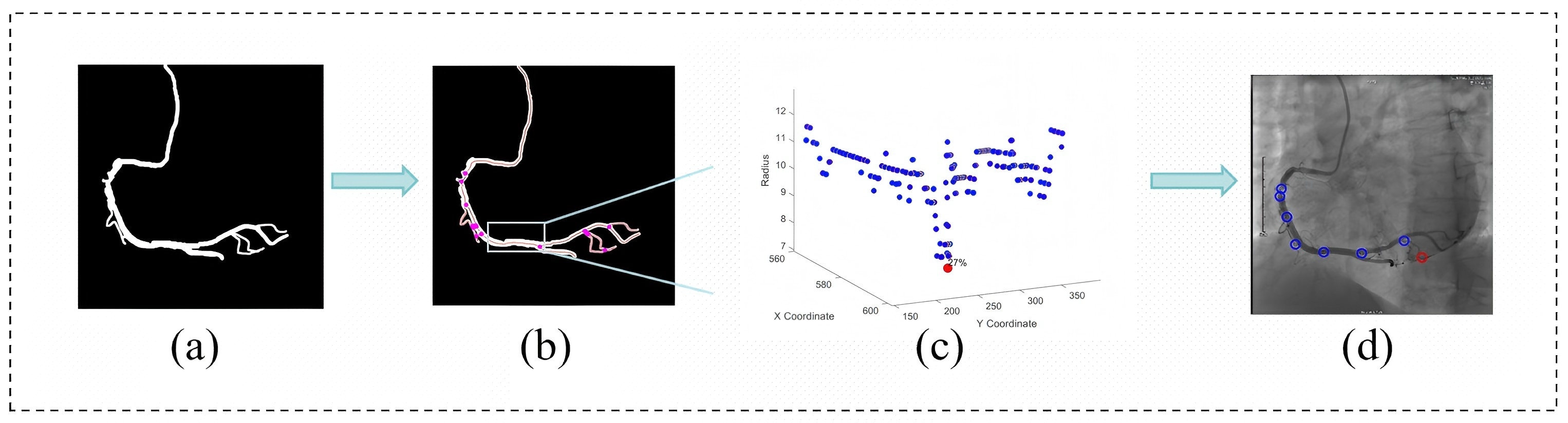} 
    \caption{Diagram of datasets for SAM-VMNet training and evaluation.}
    \label{fig2}
\end{figure*}

The workflow of our arterial segmentation and stenosis detection method is illustrated in \autoref{fig1}. Initially, angiographic images (\autoref{fig1} (a) left) were fed into our segmentation model, which employed a deep learning-based VM-UNet network enhanced with MedSAM (\autoref{fig1} (a) middle) to extract the arterial structure (\autoref{fig1} (a) right). Subsequently, various techniques were applied to identify obstructions and stenoses (\autoref{fig1} (b)).

\subsection{Image dataset}
GH dataset. In this study, we constructed an in-house dataset containing several angiographic video clips collected from the 302 Hospital of Guihang, China. The frame rates of these video clips exceed ten frames per second to ensure image continuity and the accuracy of motion capture. The dataset includes angiograms of the left and right coronary arteries, with a resolution of $800\times800$ pixels. We extracted 77 contrast-enhanced angiographic images from these video sequences, which were manually labeled by an interventional radiologist to accurately reflect the results of coronary segmentation. This labeled data provides a reliable benchmark for our study. We refer to this dataset as the GH dataset and use the Pascal VOC annotation format.

ARCADE dataset. The ARCADE dataset \cite{popov2023arcade} is specifically designed for the task of local segmentation of coronary vascular tree images and contains 1200 carefully annotated images. The dataset is originally divided into a training set (1000 images) and a validation set (200 images), with each image labeled in detail according to the Syntax Score method \cite{syntaxscore}, covering 26 different regions. However, it is important to note that some finer structures, such as the capillaries, are not annotated in this dataset.To ensure balance in our mixed dataset, we selected the 200 images from the original validation set and incorporate them into our mixed dataset.

DCA1 dataset. The DCA1 dataset \cite{cervantes2019automatic}, obtained from the department of cardiology of the Mexican Social Security Institute, comprises 141 coronary angiogram X-rays along with corresponding segmentation labels annotated by cardiologists. Each angiogram has a resolution of $300\times300$ pixels, and this dataset is mainly used to improve the overall segmentation accuracy of the model. 

Due to the limited sample size of the GH dataset, we augmented it and combined it with the ARCADE and DCA1 datasets to create a larger, more diverse training dataset of 411 images. To ensure consistency , all images in the hybrid dataset were resized to a uniform resolution of $800 \times 800$ pixels. The hybrid dataset is divided into 80\% for the training set and 20\% for the test set.\autoref{fig2} shows an example of the hybrid dataset, along with relevant information about the dataset.

\subsection{Vessel segmentation}
\subsubsection{Preliminaries}
\begin{figure}[htbp]
    \centering
    \includegraphics[width=0.6\textwidth]{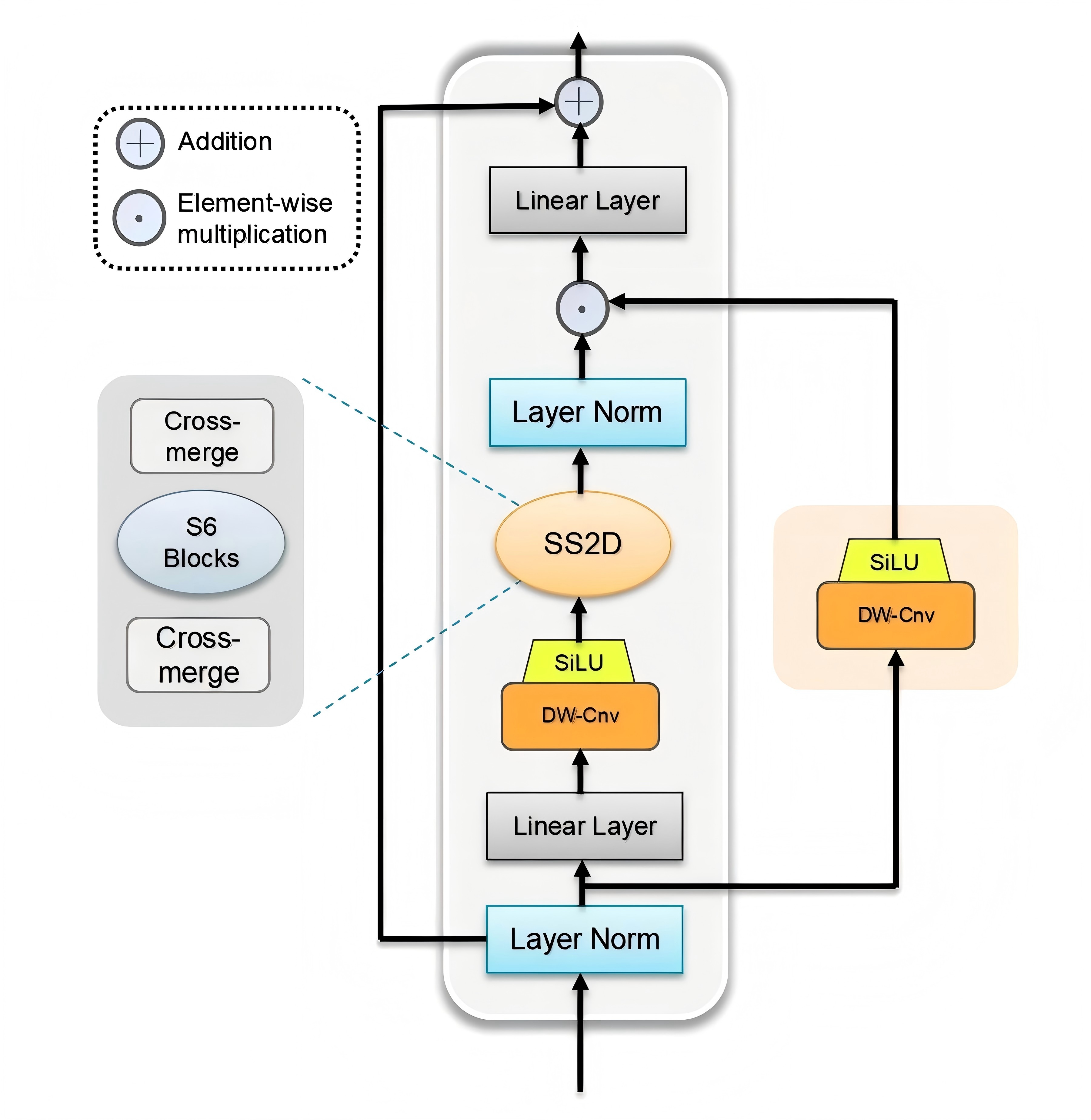}
    \caption{Structure of the vssblock.}
    \label{fig3}
\end{figure}

Vision Mamba Uet (VM-UNet) is the first medical image segmentation model based purely on state space models (SSM). The model employs an asymmetric encoder-decoder structure and introduces visual state space (VSS) blocks to capture a wide range of contextual information, resulting in excellent performance in medical image segmentation tasks. Its core VSS block is derived from VMamba, as shown in \autoref{fig3} The input is divided into two branches after layer normalization. In the first branch, the input passes through a linear layer and an activation function. In the second branch, the input is processed through a linear layer, a depth-separable convolution, and an activation function, and then fed into the 2D-selective-scan (SS2D) module for further feature extraction. Subsequently, the features are normalized using layer normalization, and the two paths are merged. Finally, the features are blended using linear layers, and this result is combined with the residual join to form the output of the VSS block.

The medical self-attention model (MedSAM) is a generalized deep learning model developed for the medical image domain. It uses a self-attention mechanism to automatically recognize important features in the input data, focuses on the relationships between different parts, and captures long-range dependencies and complex patterns. This also provides it with powerful feature extraction capabilities. The SAM network architecture consists of three main components: an image encoder, a prompt encoder, and a mask decoder. The image encoder uses a vision transformer-based model and is responsible for extracting features from input medical images. The prompt encoder processes prompts for user interaction, such as bounding boxes. Through positional encoding, the prompts are transformed into feature representations that guide the model in focusing on the region of interest. This network structure enables MedSAM to support interactive prompt-based medical image segmentation with strong feature extraction and generalization capabilities.
\begin{figure*} 
    \centering
    \includegraphics[width=\textwidth]{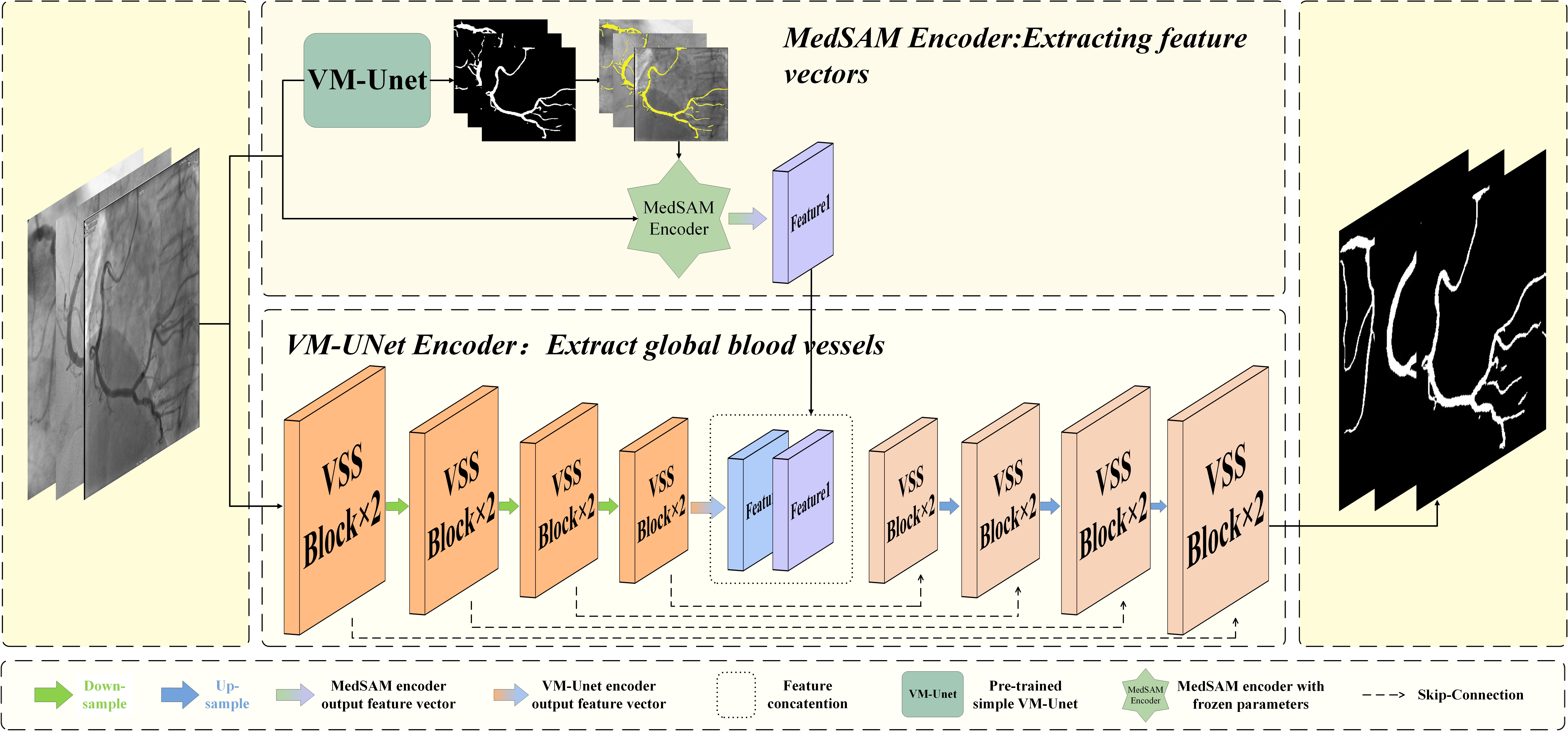}
    \caption{SAM-VMNet architecture diagram. SAM-VMNet contains two parallel branches.}
    \label{fig4}
\end{figure*}

\subsubsection{Overall architecture}
Although VM-UNet has an advantage over ViT in inference speed and expansion space, it still performs poorly when dealing with small vessels, severely stenotic vessels, or poor image quality. Inspired by nnSAM \cite{li2023nnsam}, which leverages SAM as a feature extraction tool, we aim to enhance this capability by using MedSAM's feature extraction strength and its prompt encoder's localized focus on vessels.  To this end, we designed the SAM-VMNet model, the structure of which is shown in \autoref{fig4}. It combines the technical advantages of MedSAM and VM-UNet by using two parallel encoders: the MedSAM encoder and the VM-UNet encoder. In the MedSAM encoder, we first trained a simple VM-UNet to segment the image coarsely and subsequently selected 10 points at equal intervals on its mask image, which were mapped to the original image and used as inputs to the MedSAM prompt encoder. Thereby, the feature vectors output from the MedSAM encoder are automatically acquired. In the VM-UNet encoder, the original image is fed into the four-layer stacked VSS block, and the resulting feature vectors are linearly summed with the feature vectors from MedSAM to achieve feature fusion. Finally, the output mask image is generated by the decoder of VM-UNet. MedSAM, as a large model after pre-training, serves the purpose of extracting local features, so its parameters should remain frozen during the training process to ensure the validity of the feature extraction. The weights of VM-UNet are continuously updated during training.

\subsubsection{Parallel network architecture}
By adding specific prompts to MedSAM’s input, the model can better understand the objectives of the task.      For example, hints can indicate the direction of blood vessels, emphasize the details of the vessels, and help it focus on relevant features.      Effective prompts can also speed up the model's convergence, improve the learning effect with a small number of samples, and reduce dependence on a large amount of labeled data.      SAM supports three segmentation modes: full-image segmentation mode, bounding-box mode, and point mode.      Compared with the first two, point mode can more accurately label the vessel paths and their small branches while requiring fewer computational resources, which is why this paper adopts point mode.      To automate the acquisition of prompts and reduce labor costs, we first trained a simple VM-UNet to coarsely segment the image and obtain vascular pathway prompts applicable to MedSAM.

Specifically, the image input is passed through the VM-UNet to obtain the coarsely segmented mask image.      At this point, the segmented image still contains many tiny disconnected blood vessels, but the direction and distribution of the blood vessels are roughly correct.       After conducting trials with varying point selections, 10 points in the image are chosen at equal spacing to serve as the prompt for MedSAM.      Thus, we obtained the feature vector $x_1 \in \mathbb{R}^{ 256 \times 64 \times 64}$, which emphasizes the local features of the blood vessels compared to the unprompted feature vectors.      The second encoder of SAM-VMNet is our benchmark model VM-UNet, which is based on a new visual state space block architecture.      The ability to efficiently establish long-distance dependencies while maintaining linear complexity significantly improves inference speed, making it capable of deployment on medical devices.      The image input is passed through a VM-UNet encoder consisting of four layers of VSS Block blocks to obtain the feature vector  $x_2 \in \mathbb{R}^{8 \times 8 \times 768}$.
At this point, in order to solve the problem of unclear local segmentation of blood vessels and broken connections, we would like to fuse the feature vector $x_1$ extracted by MedSAM.   By adjusting the dimensional order of \( x_1 \), applying a \( 1 \times 1 \) convolution, and using an average pooling layer, we obtain a feature vector of dimensions \( 8 \times 8 \times 768 \), matching the size of \( x_2 \). This allows us to effectively fuse the two feature vectors. The fusion process is performed using element-wise addition, which increases the information content, mitigates the problem of gradient vanishing, and reduces computational cost compared to concatenation.

This method leverages MedSAM’s prompt encoder to generate high-quality feature vectors. The fused vectors are then upsampled by the VM-UNet decoder, which outputs the final segmentation predictions. During backpropagation training, only the weights of the lower VM-UNet are updated, while the weights of MedSAM and the simple VM-UNet remain frozen, serving only for feature extraction and coarse segmentation. MedSAM’s point-based prompt method enabled precise identification of vascular paths, crucial for handling the intricate details of coronary vessels. VM-UNet, with its state-space model, effectively captures broader contextual information across the image, complementing MedSAM. This parallel architecture balances the strengths of both models, enhancing segmentation accuracy and robustness, especially in complex cases with overlapping structures or uneven contrast.
\begin{figure*}
    \centering
    \includegraphics[width=\textwidth]{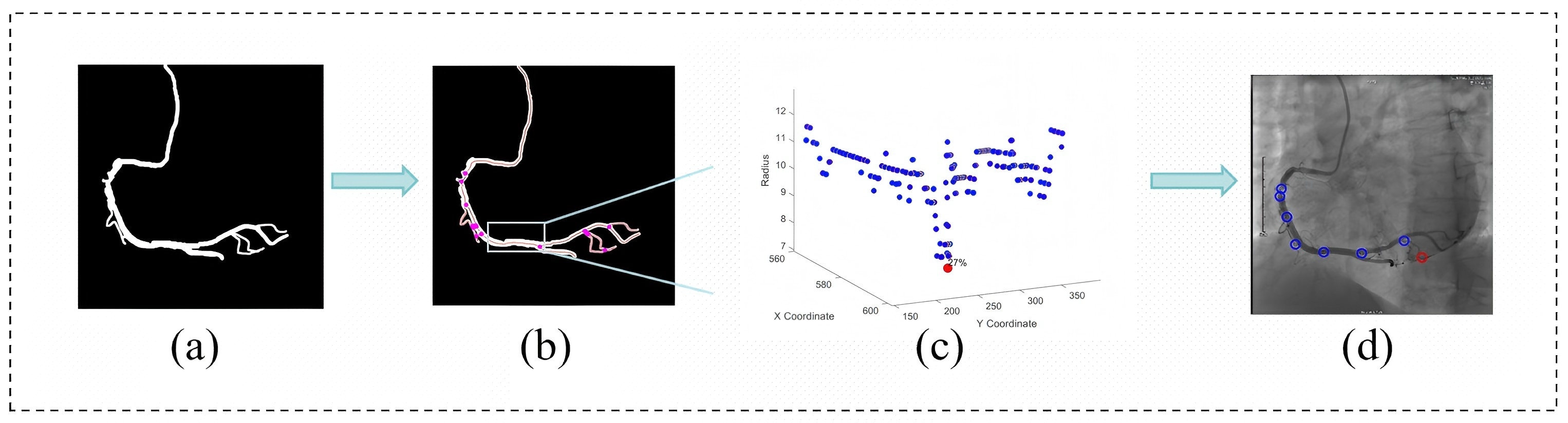}
    \caption{Stenosis detection flow chart. (a): segmentation results, (b): extraction of center line and segmentation, (c): measurement of diameter and detection of stenosis points (shown as one section), where blue points are on the center line, and red points are stenosis points,(d): final stenosis detection results,where red color
is labeled as severe stenosis, green color is moderate stenosis,
and blue color is mild stenosis.}
    \label{fig5}
\end{figure*}

\subsubsection{Loss Function}

SAM-VMNet employs a loss function that combines Binary Cross-Entropy (BCE) and Dice Loss, referred to as BceDice loss. 

The BceDice loss for two-class segmentation is defined as
\begin{equation}
L_{\text{BceDice}} = \lambda_1 L_{\text{Bce}} + \lambda_2 L_{\text{Dice}},
\end{equation}
where $\lambda_1$ and $\lambda_2$ are weighting coefficients, a typical setting is to use \( \lambda_1 = \lambda_2 = 1 \).
 $L_{\text{Bce}}$ and $L_{\text{Dice}}$ denote the Binary Cross-Entropy loss and Dice loss, respectively. The definitions of $L_{\text{Bce}}$ and $L_{\text{Dice}}$ are 
\begin{equation}
L_{\text{Bce}} = - \frac{1}{N} \sum_{i=1}^{N} \left[ y_i \log (\hat{y}_i) + (1 - y_i) \log (1 - \hat{y}_i) \right],
\end{equation}
and
\begin{equation}
L_{\text{Dice}} = 1 - \frac{2 | X \cap Y |}{| X | + | Y |},
\end{equation}
where 
 \( N \) is the total number of samples, \( y_i \) and \( \hat{y}_i \) represent the ground truth label and the predicted label for sample \( i \), respectively,  $X$ and $Y$ represent the ground truth set and the predicted set, respectively, and $|\cdot|$ denotes the cardinality of a set.

\subsection{Vascular stenosis detection}
After extracting the centerline of the coronary artery, the stenosis detection needs to be carried out. The flow chart of the entire detection process is shown in \autoref{fig5}.
\subsubsection{Arterial centerline extraction and segmentation}
To extract the centerline skeleton of the blood vessels, we used  the \texttt{bwmorph} function with the\texttt{ "thin" }operation in MATLAB. This method relies on morphological thinning, iteratively removing boundary pixels until the object reduces to minimal connected lines. By setting the third parameter to \texttt{Inf}, the operation continues until the image no longer changes \cite{guo1989parallel, lam1992thinning}. The thinning operation is defined as
\[
I_{thin} = I \setminus \bigcup_{n=1}^{M} B_n,
\]
where \( I_{thin} \) is the thinned skeleton image, \( B_n \) represents the boundary pixels removed in each erosion step, and \( M \) is the number of iterations. When set to \texttt{Inf}, the iteration continues until no more pixels are removed. After extracting the centerline, vessel segmentation is performed. A point on the centerline is classified as a segmentation point if there are three points in its 8-neighborhood. The centerline extraction and segmentation process is illustrated in \autoref{fig5} (b).

\subsubsection{Arterial diameter calculation}
To calculate the diameter of an artery, we began at a specified center point $(x_0, y_0)$.  We then traversed outward in all directions until we reached the boundary of the vessel. Specifically, the function starts from the center point and gradually expands a circular region until it encounters a point at the edge of the circle with a pixel value not equal to 255. When this point is found, the corresponding distance is recorded as the radius  $R$. The distribution of arterial diameters is illustrated in \autoref{fig5} (d).

\subsubsection{Arterial Stenosis Detection}
Traditional stenosis detection methods often rely on evaluating the diameter of an entire vessel segment, which lacks the flexibility to capture local variations within the vessel. To enable localized detection and quantification of coronary artery stenosis, we proposed a stenosis detection algorithm based on a "dynamic queue" method.

\begin{figure}[htbp]  
    \centering
    \includegraphics[width=0.6\textwidth]{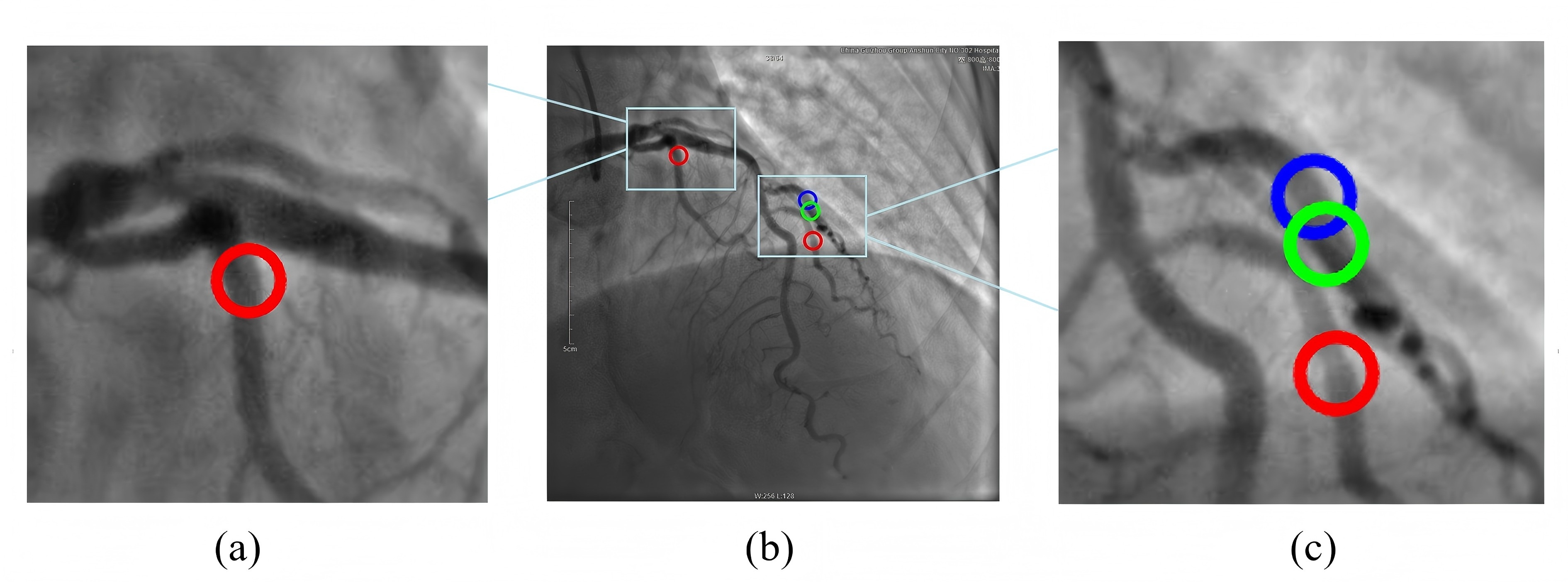}
    \caption{Schematic of arterial stenosis detection. (b) shows the results of stenosis detection in the whole vascular tree, and (a) and (c) show the local zoomed-in effect, where red color is labeled as severe stenosis, green color is moderate stenosis, and blue color is mild stenosis.}
    \label{fig6}
    
\end{figure}

This algorithm traverses each point along the path, identifies whether it is a local minimum, and calculates the corresponding degree of stenosis if it is. Specifically, during the path traversal process, when a point's radius value is smaller than that of its subsequent points, it is labeled as the potential starting point of a stenosis region. The start and end positions of the decreasing queue are recorded, while the smallest radius within the current path segment is stored as the potential stenosis point radius value. Conversely, when the radius value of a point is greater than or equal to that of its successor, it may indicate the end of the stenosis region; the termination point of the incremental queue is recorded, and the change in diameter of the stenosis segment is calculated.
The formula for calculating the degree of stenosis is 
\begin{equation}
\eta = 1 - \frac{R_c}{\frac{R_s + R_e}{2}},
\end{equation}
where $R_c$ denotes the minimum radius in the narrow region, $R_s$ and $R_e$ are the radius values at the start and end of the narrow region, respectively. The average of $(R_s + R_e)/2$ is considered the expected normal diameter at the stenosis site. By comparing $R_c$ with this average diameter, we could calculate the local degree of stenosis. This formula is used to quantify the severity of the stenosis. Ultimately, the coordinates of each stenosis point and its corresponding degree of stenosis are output. The stenosis points are graded according to the doctor's recommendation: 25\%-50\% for mild stenosis, 50\%-75\% for moderate stenosis, and $>75\%$ for severe stenosis. In addition, vessel segments with an average diameter of less than 4 pixels were excluded to avoid stenosis assessment in capillary segments. To avoid excessive clustering of stenosis points that may obscure the vascular structure, a distance threshold of $\tau=8$ was set; if the distance between adjacent stenosis points is less than $\tau$, only one stenosis point is retained. An example of arterial stenosis detection is shown in \autoref{fig6}. The entire stenosis detection algorithm is described in \hyperref[alg:stenosis]{Algorithm \ref*{alg:stenosis}}.

\begin{algorithm}[htbp] 
\caption{Stenosis detection algorithm}\label{alg:stenosis} 
\begin{algorithmic}
\State \textbf{Input:} 
\State \hspace{1em} $(x_0, y_0)$: center coordinates
\State \hspace{1em} image$(x, y)$: pixel value of the image at position $(x, y)$
\State \textbf{Output:} Coordinates of stenosis points and stenosis degree
\State \textbf{Steps:}
\State 1. Extract the centerline.
\State 2. Segment the vessel.
\State 3. \textbf{Circular search initialization:} Start from radius $r = 1$, gradually increasing to max radius $r = rr$. 
\For{each radius $r$}
    \For{each angle $\alpha$ on the circumference}
        \State Compute the pixel position and check if it meets the specified condition.
    \EndFor
\EndFor
\State 4. \textbf{Find the radius:} When the first valid pixel is found, record current radius $r$ as $R_c$.
\If{no valid pixel is found}
    \State Output the current $R_c$ as the radius.
\EndIf
\State 5. \textbf{Initialize increasing and decreasing queues:} Define the start and end points of both queues.
\State 6. \textbf{Detect vessel stenosis:}
\For{each pair of adjacent points}
    \State Extract their radius values.
    \If{current radius $>$ next radius}
        \State Mark the start of stenosis region, record start of decreasing queue, update min radius.
        \State Assign $R_s$ as the radius at the start of the stenosis region.
    \ElsIf{current radius $\leq$ next radius and decreasing queue recorded}
        \State Mark the end of stenosis region, record end of increasing queue.
        \State Assign $R_e$ as the radius at the end of the stenosis region.
    \EndIf
\EndFor
\State 7. \textbf{Calculate stenosis degree:} 

\State Calculate: $\eta =1- \frac{R_c}{\frac{R_s + R_e}{2}}.$

\State 8. \textbf{Output results:} Output the coordinates and stenosis degree for each stenosis point.
\end{algorithmic}
\end{algorithm}

\section{Experiments and results}
\subsection{Evaluation of arterial segmentation results}
For the selection of evaluation metrics, we adopted common image segmentation evaluation criteria, including IoU, F1, Accuracy, Specificity, and Sensitivity, to comprehensively measure the model's segmentation performance. 

Let TP, FP, FN, and TN denote correctly segmented target pixels, incorrectly segmented target pixels, missed target pixels, and correctly identified background pixels, respectively. The IoU defined by
\begin{equation}
\mathrm{IoU} = \frac{\mathrm{TP}}{\mathrm{TP} + \mathrm{FP} + \mathrm{FN}},
\end{equation}
is a common metric for evaluating image segmentation tasks. Higher IoU indicates better segmentation performance. The accuracy (Acc) defined by
\begin{equation}
\mathrm{Acc} = \frac{\mathrm{TP} + \mathrm{TN}}{\mathrm{TP} + \mathrm{TN} + \mathrm{FP} + \mathrm{FN}},
\end{equation}
reflects the model’s ability to correctly identify all categories. The specificity (Spe) defined by
\begin{equation}
\mathrm{Spe} = \frac{\mathrm{TN}}{\mathrm{TN} + \mathrm{FP}},
\end{equation}
indicates how well the model avoids false positives. The sensitivity (Sen) defined by \begin{equation}
\mathrm{Sen} = \frac{\mathrm{TP}}{\mathrm{TP} + \mathrm{FN}},
\end{equation} 
reflects the model’s ability to capture all positive samples. The F1 score assesses the similarity between the predicted and actual results. It is calculated as the harmonic mean of precision and recall
\begin{equation}
\mathrm{F1} = \frac{2 \mathrm{TP}}{2 \mathrm{TP} + \mathrm{FP} + \mathrm{FN}}.
\end{equation}

\subsection{Evaluation of stenosis detection}
To quantitatively assess the performance of the stenosis detection algorithm, we must first clarify several definitions. The stenosis regions in the vessel images are annotated based on the physician's diagnostic records. A stenosis point is classified as a true positive (TP) sample if it is successfully detected in both the predicted arterial contour and the true annotation. If a stenosis is present in the true annotation but not detected in the prediction, it is classified as a false negative (FN) sample. Conversely, if a stenosis is detected in the predicted arterial profile without a corresponding stenosis in the actual labeling, it is classified as a false positive (FP) sample.

\begin{figure} [H]
    \centering
    \includegraphics[width=0.8\textwidth]{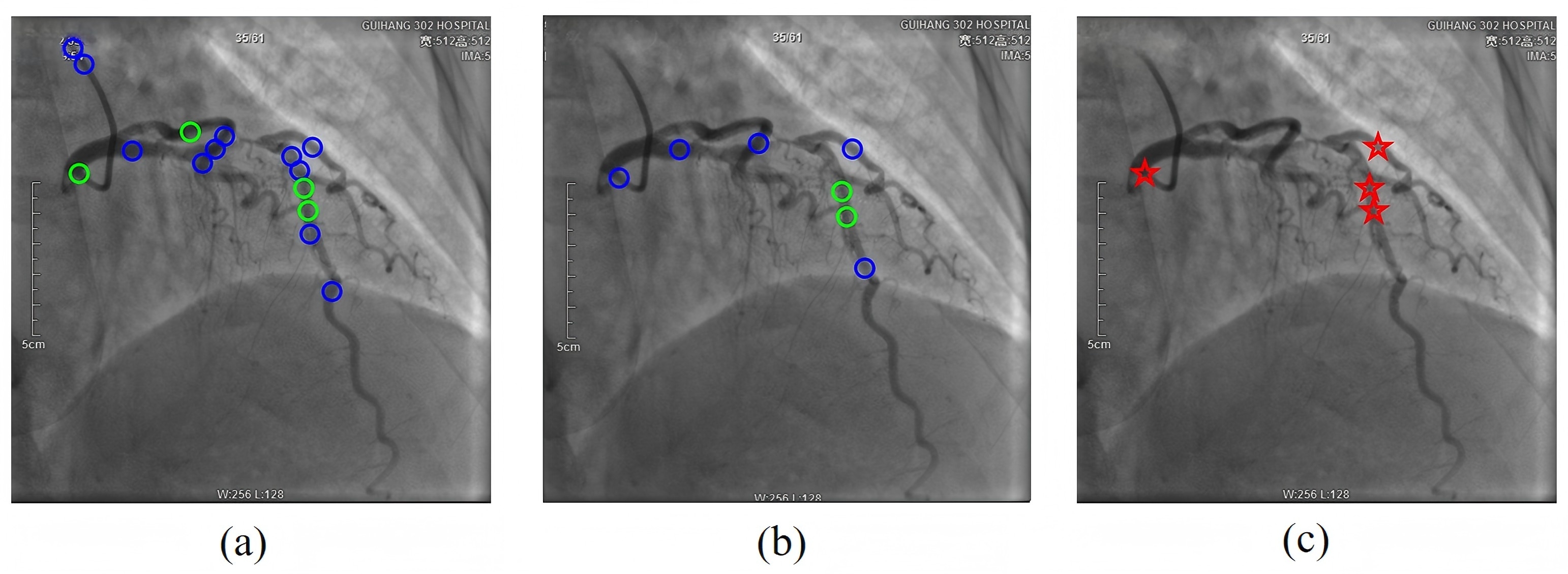}
    \caption{Illustration of the rules for the evaluation of stenosis detection algorithm.  (a) The stenosis points labeled based on the doctor’s diagnostic records.
(b) The detected stenosis points using the arterial contours from the model prediction.  (c) The matching results of the detected stenotic points. }
    \label{fig7}
\end{figure}

A "match" between a predicted stenosis and a labeled stenosis is defined as having a distance less than a specified threshold, \( \gamma = 10 \). The matching results are illustrated in \autoref{fig7}.

Similar to the evaluation of a binary classification task, we used the true positive rate (TPR) 
\begin{equation}
\mathrm{TPR} = \frac{\mathrm{TP}}{\mathrm{TP} + \mathrm{FN}},
\label{eq:tpr}
\end{equation}
and the positive predictive value (PPV)
\begin{equation}
\mathrm{PPV} = \frac{\mathrm{TP}}{\mathrm{TP} + \mathrm{FP}},
\label{eq:ppv}
\end{equation}
to assess the performance of the stenosis detection algorithm. In addition, we evaluated the results of stenosis detection using the absolute root mean square error (ARMSE) 
\begin{equation}
\mathrm{ARMSE} = \sqrt{ \frac{1}{M} \sum_{1}^{M} \left(n_f - n_l \right)^2 },
\label{eq:armse} 
\end{equation}
and the relative root mean square error (RRMSE) 
\begin{equation}
\mathrm{RRMSE} = \sqrt{ \frac{1}{M} \sum_{1}^{M} \left( \frac{n_f - n_l}{n_l} \right)^2 },
\label{eq:rrmse}  
\end{equation}
where\( M \)is the total number of samples,$n_f$ denotes the number of stenosis points for stenosis detection based on the predicted arterial contour, and $n_l$ is the number of stenosis points labeled by the corresponding physician.

\subsection{Arterial segmentation results}

To assess the segmentation performance of SAM-VMNet on coronary angiography images, we compared it with several widely recognized deep learning networks for medical image segmentation. These include UNet \cite{ronneberger2015unet}, UNet++ \cite{zhou2018unet++}, TransUNet \cite{chen2021transunet}, MALUNet \cite{ruan2022malunet}, Transfuse \cite{zhang2021transfuse}, and MISSformer \cite{huang2023missformer}. We also included the original VM-UNet \cite{ruan2024vm} for comparative analysis. All networks were trained on the same mixed dataset for 200 epochs. The quantitative comparison of these models is presented in Table \ref{tab:model_comparison}.

\begin{figure} [H]
    \centering
    \includegraphics[width=\textwidth]{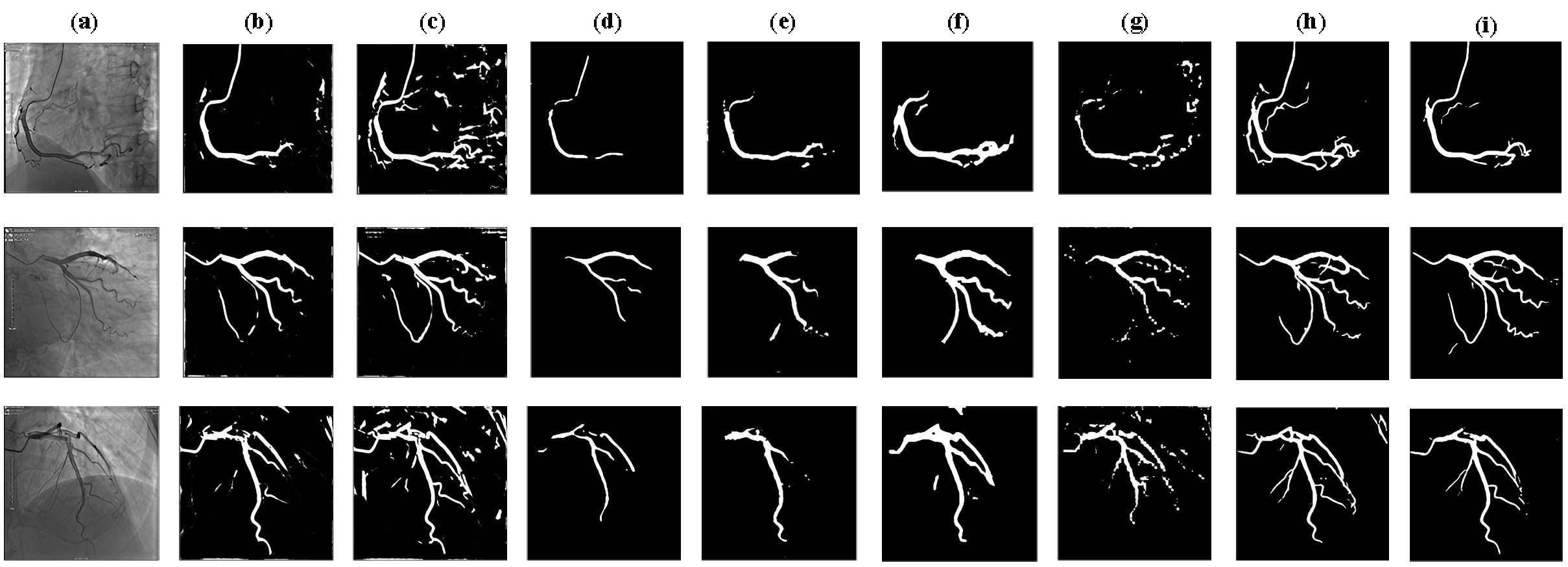}
    \caption{Illustration of the artery segmentation results from right coronary artery (top row) and left coronary artery (bottom two rows), obtained by (b) UNet, (c) UNet++, (d) TransUNet, (e) MALUNet, (f) Transfuse, (g) MISSformer, (h) VM-UNet, and (i) SAM-VMUNet. The original images are shown in (a).}
    \label{fig8}
\end{figure}

An example of the resulting arterial contour is shown in \autoref{fig8}. We also compared the ARCADE dataset with the full public dataset, with the quantitative comparison metrics displayed in Table \ref{tab:model_comparison2}. As indicated in Table \ref{tab:model_comparison}, SAM-VMNet outperformed other models in the IoU,Acc,Spe and F1 evaluation metrics on the hybrid dataset, demonstrating that our proposed network is significantly competitive in these aspects. According to Table \ref{tab:model_comparison2}, in the ARCADE dataset, SAM-VMNet also excels in these evaluation metrics. These experimental results further confirmed the superior performance of the SSM model in the field of medical image segmentation.

\begin{table}[htbp]
    \centering
    \caption{Comparison of different segmentation models for IoU, Acc, Spe, Sen, F1 on the mixed dataset}
    \label{tab:model_comparison}
    \setlength{\tabcolsep}{2pt}  
    \begin{tabular}{lccccc}
        \toprule
        Model & IoU & Acc & Spe & Sen & F1 \\
           \midrule

       UNet\cite{ronneberger2015unet}
             & 0.6307  & 0.9746 & 0.9870   &  0.7456 & 0.7523 \\
        UNet++\cite{zhou2018unet++}            & 0.3836  & 0.9548 & 0.8109  & 0.4213 & 0.5545 \\
        Transunet\cite{chen2021transunet}         & 0.3987  & 0.9546 & 0.7651  & 0.4781 & 0.5426 \\
        MALUNet\cite{ruan2022malunet},  & 0.5576  & 0.9705 & 0.9851  & 0.7076 & 0.7160 \\
        Transfuse\cite{zhang2021transfuse}         & 0.5709  & 0.9552 & 0.9681  & \textbf{0.8106} & 0.7249 \\
        Missformer\cite{huang2023missformer}       & 0.4072  & 0.9452 & 0.6706  & 0.5126 & 0.5741 \\
        VM-UNet\cite{ruan2024vm}           & 0.5445  & 0.9765 & 0.9902  & 0.7293 & 0.7658 \\
        SAM-VMNet  &  \textbf{0.6308}  & \textbf{0.9772} & \textbf{0.9903}  & 0.7409 &  \textbf{0.7736} \\
        \bottomrule
    \end{tabular}
\end{table}

\begin{table}[H]
    \centering
    \caption{Comparison of different segmentation models for IoU, Acc, Spe, Sen, F1 on the ARCADE dataset}
    \label{tab:model_comparison2}
    \setlength{\tabcolsep}{2pt}  
    \begin{tabular}{lccccc}
        \toprule
        Method        & IoU   & Acc    & Spe    & Sen    & F1 \\
        \midrule
        UNet\cite{ronneberger2015unet}          & 0.5153 & 0.9656 & 0.5226 & 0.5325 & 0.5523 \\
        UNet++\cite{zhou2018unet++}         & 0.5856 & 0.9808 & 0.7647 & 0.7143 & 0.7387 \\
        TransUNet\cite{chen2021transunet}      & 0.4364 & 0.9762 & 0.7569 & 0.5141 & 0.5889 \\
        MALUNet\cite{ruan2022malunet}       & 0.5561 & 0.9801 & 0.9915 & 0.6801 & 0.7147 \\
        Transfuse\cite{chen2021transunet}     & 0.5737 & 0.9769 & 0.9819 & \textbf{0.8511} & 0.7210 \\
        Missformer\cite{huang2023missformer}    & 0.2381 & 0.9693 & 0.6863 & 0.2704 & 0.3680 \\
        VM-UNet\cite{ruan2024vm}        & 0.5445 & 0.9773 & 0.9886 & 0.6987 & 0.7051 \\
        SAM-VMNet     & \textbf{0.6303} & \textbf{0.9832} & \textbf{0.9933} & 0.7343 & \textbf{0.7733} \\
        \bottomrule
    \end{tabular}
\end{table}

\subsection{Stenosis detection results}
To quantitatively demonstrate the performance of the stenosis detection algorithms, we evaluated the true positive rate (TPR) and positive predictive value (PPV) defined in  Eqs.~\eqref{eq:tpr} and \eqref{eq:ppv}, using the images annotated with stenotic regions as the reference standard. For our test set, we selected a right coronary angiogram image and 1-2 left coronary images for each patient, resulting in a total of 209 annotated coronary angiogram images with varying degrees of stenosis after initial screening.

\textcolor{blue}{Table \ref{tab:tpr_ppv_comparison}} presents the results of the quantitative detection of different types of stenosis using arterial contouring techniques, including UNet, UNet++, Transunet, MALUNet, Transfuse, MISSformer, and VM-UNet. As shown in \textcolor{blue}{Table \ref{tab:tpr_ppv_comparison}}, the stenosis detection algorithm achieves a TPR of 0.5332 and a PPV of 0.6032 using the contours generated by the proposed SAM-VMNet.
\begin{table}[htbp]
    \centering
    \caption{Stenosis detection results of different models.}
    \label{tab:tpr_ppv_comparison}
    \setlength{\tabcolsep}{4pt}  
    \begin{tabular}{lcccc}
        \toprule
        Model & TPR & PPV & ARMSE & RRMSE \\
        \midrule
        UNet\cite{ronneberger2015unet}          & 0.5567 & 0.5616 & 1.1667 & 0.2821 \\
        UNet++\cite{zhou2018unet++}        & 0.5833 & 0.5714 & 1.2583 & 0.2143 \\
        Transunet\cite{chen2021transunet}     & 0.5911 & 0.5839 & 1.6543 & 0.2054 \\
        MALUNet\cite{ruan2022malunet}       & \textbf{0.6283} & 0.5725 & 1.3133 & 0.4844 \\
        Transfuse\cite{zhang2021transfuse}     & 0.5945 & \textbf{0.5933} & 1.3657 & 0.3914 \\
        Missformer\cite{huang2023missformer}    & 0.5732 & 0.5524 & 1.7068 & 0.1763 \\
        VM-UNet\cite{ruan2024vm}       & 0.5736 & 0.5734 & 1.7382 & 0.1915 \\
        SAM-VMNet      & 0.5867 & 0.5911 & \textbf{1.1667} & \textbf{0.1673} \\
        \bottomrule
    \end{tabular}
\end{table}

\section{Discussion}
\subsection{Performance analysis of arterial segmentation}
\begin{figure*} 
    \centering
   \includegraphics[width=\textwidth]{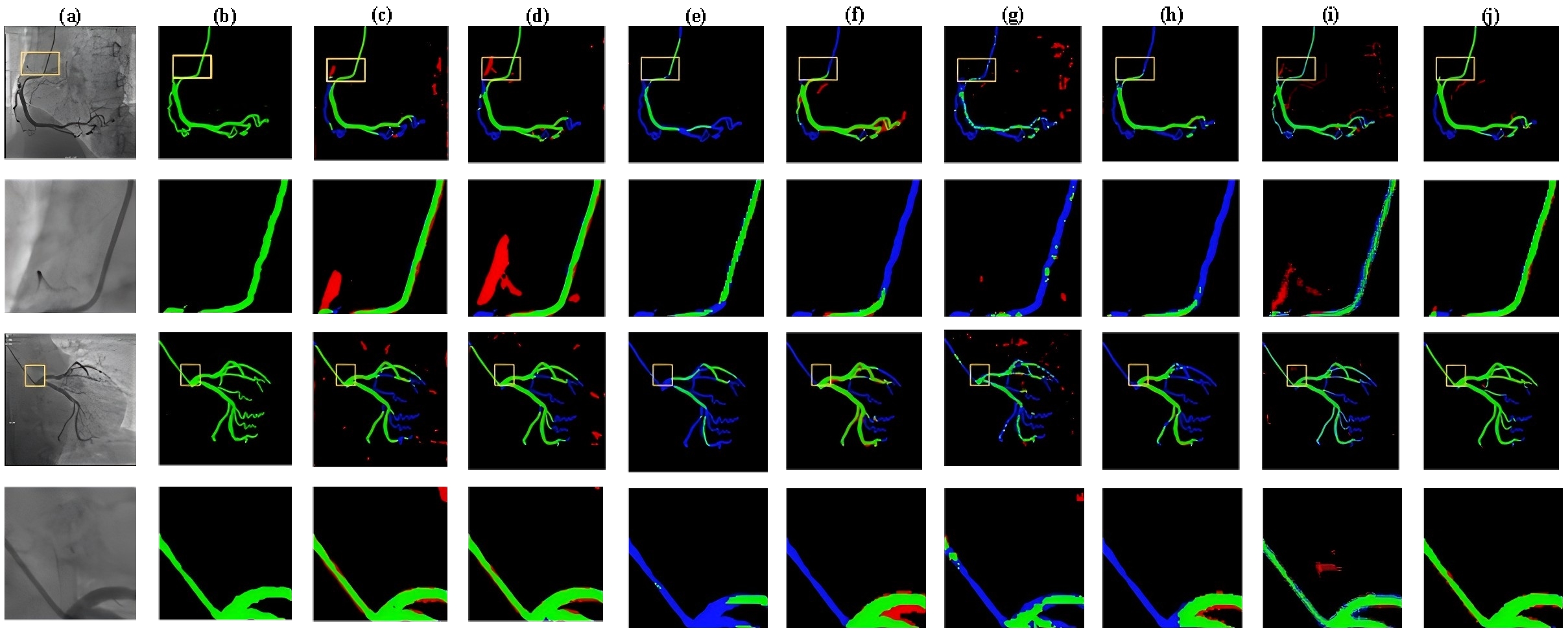}
    \caption{Difference map between segmentation results and ground truth. Green, red, and blue regions indicate true positive, false negative, and false positive pixels, respectively. The regions of interest have been magnified accordingly. (a) Original images; (b) ground truth; (c) UNet; (d) UNet++; (e) TransUNet; (f) Transfuse; (g) MISSFormer; (h) MALUNet; (i) VM-UNet; (j) SAM-VMUNet.}
    \label{fig9}
\end{figure*}

To further illustrate the effectiveness of the proposed SAM-VMNet, we compared its segmentation results and the difference maps of the contrasting models with the actual situation, as shown in \autoref{fig9}. A true positive pixel indicates that the pixel belongs to an artery and is correctly predicted by our model as an arterial pixel. A false positive pixel is a background pixel that is incorrectly predicted as an arterial pixel. A false negative pixel is an arterial pixel that is incorrectly predicted as a background pixel. As shown in \autoref{fig9}, SAM-VMNet had fewer false positives and false negatives than the other methods, with vessels exhibiting stronger connectivity and being less affected by background noise and artifacts. 

On the mixed dataset, as shown in \textcolor{blue}{Table \ref{tab:model_comparison}}, SAM-VMNet achieved an IoU of 63.08\%, accuracy of 97.72\%, specificity of 99.03\%, and an F1 score of 77.36\%, which are the highest among all models. The high IoU score indicates that SAM-VMNet could effectively capture the overlap between predicted vessel regions and the ground truth, ensuring precise segmentation boundaries and minimizing segmentation errors. Additionally, SAM-VMNet’s sensitivity score reached 0.7409, representing an improvement of 1.18\% over the baseline model, VM-UNet. This indicated that SAM-VMNet can more comprehensively detect vessel regions, thereby reducing missed detections. The F1 score also increased from 0.7658 to 0.7736, further validating SAM-VMNet’s overall advantage in vascular segmentation tasks. These results demonstrated that SAM-VMNet is more accurate in detecting vessel regions, which is crucial for downstream tasks that require finer vessel segmentation.

On the ARCADE dataset, more data (1,000 sheets) were added for training, and according to \textcolor{blue}{Table \ref{tab:model_comparison2}}, SAM-VMNet shows even better performance, achieving optimal scores in IoU, accuracy, specificity, and F1 metrics. This suggests that our proposed model has greater potential for handling larger datasets. Overall, this analysis indicates that SAM-VMNet exhibits stronger robustness and accuracy in the vascular segmentation domain and can effectively provide the necessary technical support for stenosis detection in downstream tasks.
\begin{figure}
    \centering
    \includegraphics[width=0.8\textwidth]{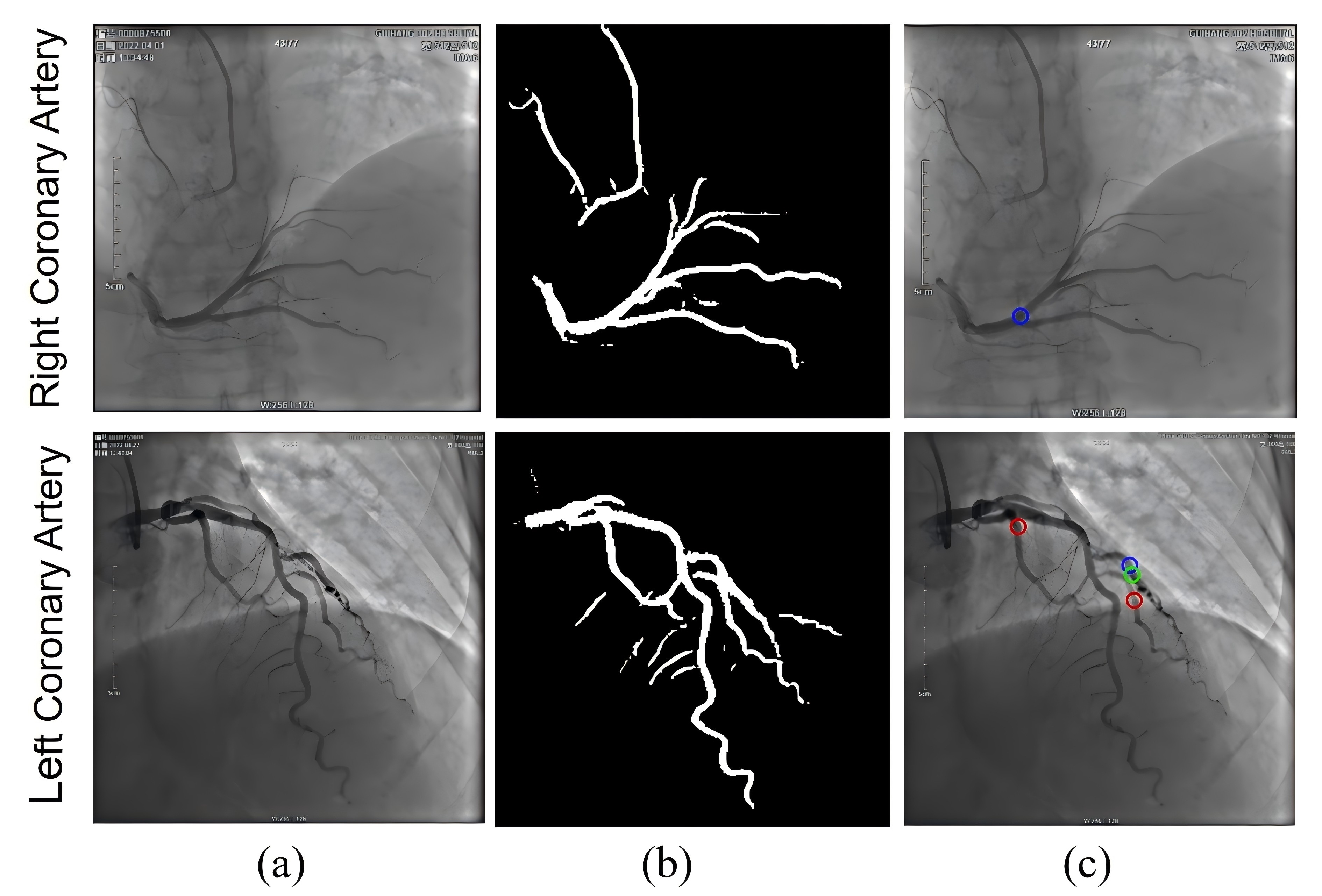}
    \caption{Example results of the stenosis detection algorithm. (a) Original image. (b) Artery contour segmentation results. (c) Stenosis detection results using SAM-VMNet.}
    \label{fig10}
\end{figure}
\subsection{Performance analysis for stenosis detection}

\textcolor{blue}{Table\ref{tab:tpr_ppv_comparison}} indicates that stenosis detection using SAM-VMNet for extracting vessel contours demonstrates significant advantages in several key metrics, particularly in RRMSE, which achieved an impressive minimum value of 0.1673.  This result highlights our model's robustness and generalization ability in handling various levels of stenosis, delivering stable and accurate predictions.  Additionally, the model excelled in PPV and TPR, achieving values of 0.5911 and 0.5867, respectively.  This high accuracy signifies that the model is reliable in predicting stenotic regions, effectively reducing false alarms, which is crucial for clinical diagnosis and decision-making.  Although the ARMSE value was 1.7338, which was not optimal among all models, it remained at a sufficiently high level, ensuring effective identification of real lesions and reasonable overall prediction bias.  Overall, our model demonstrated superior performance in stenosis detection, particularly in reducing false alarms and enhancing predictive stability, which is vital for improving the accuracy and reliability of clinical diagnoses.\autoref{fig10} shows example results of the stenosis detection algorithm.

\subsection{Clinical Overview and Applications}
The anatomic depiction of coronary lesions by , coronary angiography remains the gold standard for diagnosing coronary artery stenosis.   However, an important limitation of coronary angiography is its dependence on visual assessment to determine the degree of stenosis.   This qualitative rather than quantitative estimation of stenosis has been questioned in recent years because of significant variability between observers.  We proposed an automated coronary artery identification model and developed an algorithm to detect stenosis in arterial segments.  Utilizing an end-to-end detection method without any data pre-processing will streamline the assessment of coronary stenoses, particularly in cases of diffuse stenosis or serial angiograms, where visual estimation can often lead to an underestimation of disease severity.
\subsection{Limitations}
There are several limitations to the proposed methods.  First, the memory of the SSM is inherently lossy, which makes it less effective than the lossless memory of the attention mechanism.  Mamba struggles to demonstrate its advantages when processing short sequences, potentially resulting in the loss of local features or small structural elements in the image—areas where the attention mechanism excels.

Second, regarding cue point selection, although our approach using a pre-trained VM-UNet for rough segmentation and cue point generation performed well in our study, its effectiveness may be compromised by the quality of the initial segmentation results.  If these initial results are not sufficiently accurate, the subsequent MedSAM feature extraction and, consequently, the final segmentation results may be negatively impacted.

Furthermore, while our method shows promise in the coronary vascular dataset, its ability to generalize to other medical image segmentation tasks requires further verification.  Different types of medical images present unique characteristics and challenges.  Therefore, a broader range of medical image datasets is necessary to thoroughly validate the generalization and stability of SAM-VMNet.

Additionally, the sample size for this study is quite limited, as the public coronary angiography dataset is small and manual labeling is time-consuming.  More images are needed to enhance the accuracy of both training and testing.  Moreover, our method employs a 2D detection approach, which may not accurately capture the true 3D structure of the vessels, leading to potential measurement errors in vessel shape and size.

\section{Conclusion}
In this paper, we proposed two key innovations: a novel coronary artery segmentation network architecture, SAM-VMNet, and a dynamic queue method for stenosis detection. SAM-VMNet seamlessly combines the robust feature extraction capabilities of MedSAM with the long-range dependency modeling of VM-UNet, significantly enhancing the segmentation quality of coronary angiography images. The dynamic queue method provides an effective tool for capturing local vascular stenosis, enabling physicians to more accurately identify areas of coronary artery narrowing, thus reducing misdiagnosis and improving diagnostic accuracy. Our method holds great potential for clinical applications, providing pre-diagnostic support to aid in the early detection and precise diagnosis of coronary artery disease.

\section*{Data availability}
Since the GH dataset used in this work contains patient data, these cannot be made generally available to the public
due to privacy concerns. However, we utilized the ARCADE dataset for external validation, which can be accessed through the link provided by the author: \url{https://zenodo.org/records/8386059}.

\section*{Code Availability}
The code for the SAM-VMNet architecture, along with examples of synthetic angiograms, is publicly accessible on GitHub at the following repository: \url{https://github.com/qimingfan10/SAM-VMNet}. This code is distributed under the Polyform Noncommercial license.

\section*{Declaration of competing interest}
The authors declare that there are no known financial interests or personal relationships that could have influenced the work presented in this paper.

\section*{Acknowledgments}
This work is supported by the Qingdao Natural Science Foundation
(No. 23-2-1-158-zyyd-jch) and the Fundamental Research Funds for the Central
Universities of China (No. 202264006).

\clearpage

\section*{References}
\addcontentsline{toc}{section}{\numberline{}References}
\vspace*{-20mm}





\bibliography{example}      



\bibliographystyle{medphy.bst}    


\end{document}